\documentstyle[aps,pre,multicol,epsf]{revtex}
\begin{document}
\draft
\title{Domain Growth in a 1-D Driven Diffusive System}
\author{Stephen J.\ Cornell and Alan J.\ Bray}
\address{
Department of Theoretical Physics,\\
 University of Manchester\\
Manchester M13 9PL, UK.
}
\date{March 21, 1996}
\maketitle
\begin{abstract}
The low-temperature coarsening dynamics of a one-dimensional Ising model, 
with conserved magnetisation and subject to a small external driving force, 
is studied analytically in the limit where the volume fraction $\mu$ of the 
minority phase is small, and numerically for general $\mu$. The mean domain 
size $L(t)$ grows as $t^{1/2}$ in all cases, and the domain-size 
distribution for domains of one sign is very well described by the form 
$P_l(l) \propto (l/L^3)\exp[-\lambda(\mu)(l^2/L^2)]$, which is exact 
for small $\mu$ (and possibly for all $\mu$). The persistence exponent 
for the minority phase has the value $3/2$ for $\mu \to 0$. 
 
\end{abstract}

\begin{multicols}{2}
\section{Introduction}
The field of phase-ordering dynamics is by now quite well developed 
\cite{Review}. It deals with the approach to equilibrium of a system 
quenched from a homogeneous high-temperature phase into a two-phase region. 
Familiar examples are binary alloys and binary liquids, which are described 
by a scalar order parameter. Recent work has addressed also cases where 
the order parameter symmetry is continuous rather than discrete \cite{Review}. 
Especially interesting is the scaling regime that emerges in the late stages 
of growth. For a scalar order parameter, for example, the domain morphology 
is apparently time-independent if lengths are scaled to a single 
time-dependent length scale $L(t)$, which represents the typical `domain 
size'. This implies that two-point correlation functions depend on the spatial 
separation ${\bf r}$ of the points only through the ratio $|{\bf r}|/L(t)$. 

By contrast, the coarsening dynamics of driven systems has been much less 
studied. A physically relevant example, indeed the motivation for the present 
work, is the phase separation of a binary liquid under gravity. Numerical 
simulations of a (physically less realistic) alloy model with gravity suggest 
the existence of two growing length scales, parallel and perpendicular to the 
field \cite{Yeung} - \cite{Alex}, although it has proved difficult to 
unambiguously extract the time-dependence of these length scales. 

An independent field of study concerns the stationary properties of these 
`driven diffusive systems' \cite{Zia}. Here we focus on the nonstationary, 
coarsening dynamics which, as we have noted, has attracted relatively little 
attention thus far. Ultimately, we would like to understand the coarsening 
of binary liquids under weak gravity, including hydrodynamic effects 
\cite{Siggia}, but as a first step we settle here for a less ambitious goal. 
Specifically we study a one-dimensional Ising model with conserved dynamics 
and a driving field $E$ which favours transport of `up' spins to the right 
(and `down' spins to the left). We work in the regime $T \ll E \ll J$, where 
$T$ and $J$ are the temperature and exchange coupling respectively. We derive 
exact results in the limit where one phase occupies a small volume fraction, 
and numerical results for general volume fractions. The main results are a 
$\sqrt{t}$ dependence for the mean domain size, and a domain-size 
distribution for domains of one sign of the form 
$P_l(l) \propto (l/L^3)\exp(-\lambda l^2/L^2)$, where $L$ is the mean size 
of domains of that sign. We also show that the recently introduced 
\cite{persistence} `persistence exponent' $\theta$, which describes the 
fraction $f(t) \sim t^{-\theta}$ of 
spins of one phase which have not flipped up to time $t$ (in a sense to be 
clarified below) is $\theta = 3/2$ for the minority phase in the limit where 
that phase has a vanishingly small volume fraction. These are the first 
analytical results for the coarsening dynamics of this driven diffusive 
system.

The paper is organised as follows. In the following section we define the 
model. 
In section III we discuss domain growth and dynamical scaling in the model, 
while section IV deals with the persistence exponent. Section V concludes with 
a summary and discussion of the results. 

\section{The Model}
The microscopic model we consider is a chain of Ising spins $S_i=\pm 1$
with nearest-neighbour coupling strength $J$.  The system evolves by 
nearest-neighbour spin-exchange dynamics, with a driving force $E$ that 
favours motion of `up' spins to the right over motion to the left.  
That is, the microscopic processes are
\newcommand\ua{\uparrow}
\newcommand\da{\downarrow}
\newcommand\lra{\rightleftharpoons}
\begin{center}
\begin{tabular}{ccll}
$+$$+$$-$$-$ \ $\lra$ \ $+$$-$$+$$-$ &\qquad& $\Delta=4J-E$& (i)\\
$-$$-$$+$$+$ \ $\lra$ \ $-$$+$$-$$+$ &\qquad& $\Delta=4J+E$& (ii)\\
$+$$+$$-$$+$ \ $\lra$ \ $+$$-$$+$$+$ &\qquad& $\Delta=-E$  & (iii)\\
$-$$+$$-$$-$ \ $\lra$ \ $-$$-$$+$$-$ &\qquad& $\Delta=-E$ &  (iv), 
\end{tabular}
\end{center}
\newcommand\rh{$\rightharpoonup$}
\newcommand\lh{$\leftharpoondown$}
where the rate for a process from left to right is proportional to
$(1/2)(1-\tanh(\Delta/2T))$.  We distinguish between the `forward' and 
`backward' versions of the processes depicted above by using \rh \ and\ 
\lh \ to denote the process from left-to-right and right-to-left respectively.

We consider the regime $T\ll E\ll J$.  This is very different from other 
studies, which have concentrated on the limit $J/T\to 0$ \cite{Derrida}.  
The system possesses metastable states consisting of long domains of parallel 
spins, separated by domain walls.  After a long time of order $\exp[(4J-E)
/T]$, either one of processes (i\rh) or (ii\rh) takes place, i.e.\  
a spin splits off from a domain.  If process (ii\rh) takes place, the system
quickly relaxes back to the metastable state by the reverse 
process (ii\lh), since all the other possible processes are endothermic.
If process (i\rh) occurs, the system can relax further by exothermic 
processes of the kind (iv\rh), so that the `up' spin moves to the right, 
eventually meeting and adhering to a domain wall; the system may also relax 
exothermically by processes (iii\rh), so that a `down' spin moves left, 
eventually meeting another domain wall.  The motion of these free spins 
is unidirectional, because the reverse processes (iii\lh) and (iv\lh)
are inhibited by a factor of order $\exp(-E/T)$.  

The result of the free `up' spin moving to the right is for the `down'
domain through which the spin has travelled to have moved bodily one 
step to the left; the result of a free `down' spin moving to the left 
is for the `up' domain to the left of the wall to take one step to the 
right. We may therefore map the microscopic dynamics of the lattice of
spins onto one for an array of domains.  The system then evolves by a 
domain of `up' spins moving spontaneously to the right,  or a domain of 
`down' spins moving to the left.  The rates for such processes are 
independent of the domain size.  This mapping is analogous to a mapping 
by Huse and Majumdar \cite{HuseMajum} for the low-temperature 
Kawasaki chain, corresponding to this model in the opposite limit $E=0$.

When domains are of size two spins or less, they can vanish.  The microscopic 
mechanism for this in the Ising spin picture is series of events 
involving two random walkers that may coalesce, which translates into rather 
complex transitions in the domain representation.  However, when 
the domains are large, the details of this domain annihilation process are not 
expected to be important, so we choose to study a model where the 
simple domain-shifting  dynamics applies for domains down to size one, and 
removing a domain if its size reduces to zero.  Simulations of this 
simplified system permitted much better statistics than would be possible 
with the true, microscopic system, whilst still giving indistinguishable 
scaling behaviour.

The algorithm used for simulation was the following: (i) set up an array of
alternating `up' and `down' domains; (ii) pick a domain at random; 
(iii) if the domain is `up', move to the right [i.e., reduce the size of its 
right neighbour by one, and increase its left neighbour by 1], otherwise move 
to the left; (iv) if one of the neighbouring domains is of zero size then 
remove it, merging its neighbours; (v) update the clock by 1/({\sl number of 
domains}); (vi) repeat steps (ii)--(vi).

\section{Domain Growth and Dynamical Scaling}
\subsection{Simulation results}
Since the number of domains in the model can decrease but not increase, 
the average domain size must increase monotonically.  In a finite system 
(with periodic boundary conditions) the system will coarsen until there is 
only one `up' domain and one `down' domain; this state will not be stationary, 
because the dynamics still permits both domain walls to perform correlated 
random walks.  There will be a wide regime of time during which the average 
domain size is much smaller than the system size, and the system might be 
expected to display dynamic scaling.

Simulations using the domain model described in the previous section were 
performed, using lattice sizes in the region $10^5$--$10^6$ spins, 
for times up to $3\times 10^4$ and averaging over several hundred samples.  
Random initial conditions, where a spin has a probability $\mu$ of being 
`up' and $(1-\mu)$ of being `down', were used; similar results were found 
if an ordered initial state was prepared using alternating single `up' spins
and domains of $(1-\mu)/\mu$ `down' spins.  The dynamics conserves the 
magnetisation, so the volume fraction $\mu$ remains unchanged.  Several 
different random number generators were used, and the results checked
for consistency; the four-register shift generator of Ziff \cite{Ziff} 
was used for the runs of highest statistics, once it had been established 
that it gave results consistent with other generators.

\begin{figure}
\narrowtext
\epsfxsize=\hsize
\epsfbox{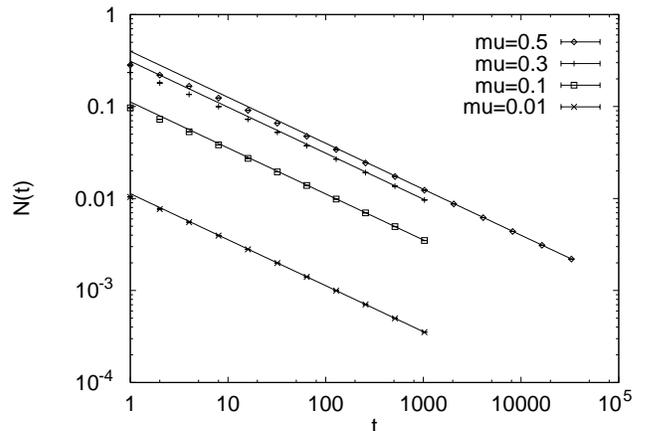}
\caption{The average domain density $N(t)$ for four values of the volume 
fraction $\mu$.  The straight lines are plots of the asymptotic prediction 
from Eqn.\ (\protect{\ref{resultN}})\label{fig1}}
\end{figure}

\begin{figure}
\epsfxsize=\hsize
\epsfbox{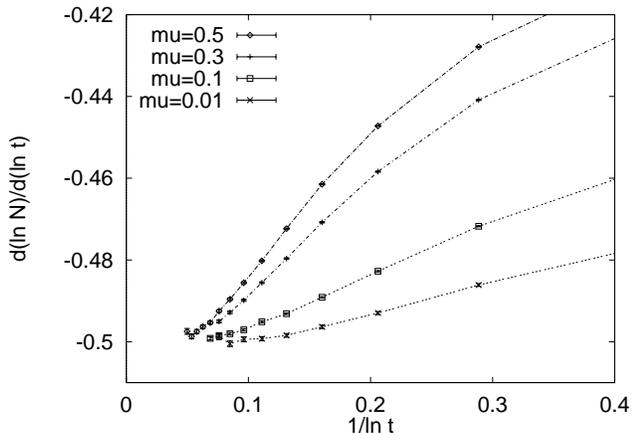}
\caption{The effective exponent for the decay of the domain wall 
density.\label{fig2}}
\end{figure}

Figure \ref{fig1} shows the average domain density plotted, for different 
volume fractions $\mu$, as a function of time $t$ on a
log-log scale.  The straight lines all have gradient $-0.50$.  
Figure \ref{fig2} shows a time-dependent effective exponent, defined as 
the gradient of a line between successive points in Fig.\ \ref{fig1}, 
plotted as a function of $1/\ln(t)$; the results show that the data appear 
to approach the value $-0.50$ as $t\to\infty$, though the convergence 
is slower for larger values of $\mu$.  The characteristic domain size 
therefore increases like $t^{0.50}$.

\begin{figure}
\epsfxsize=\hsize
\epsfbox{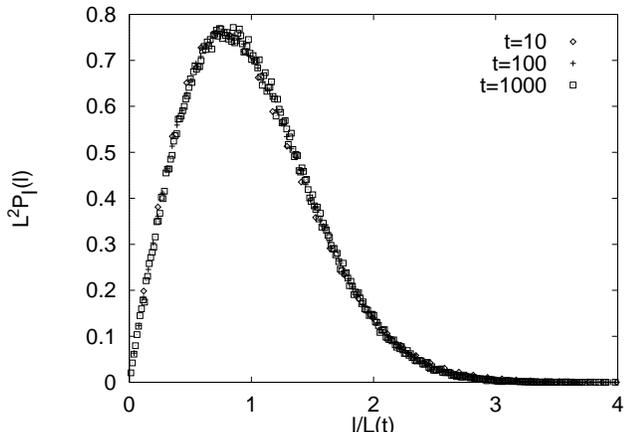}
\caption{Scaling plot of the domain size distribution for volume fraction 
$\mu=0.5$.
\label{fig3}}
\end{figure}

\begin{figure}
\epsfxsize=\hsize
\epsfbox{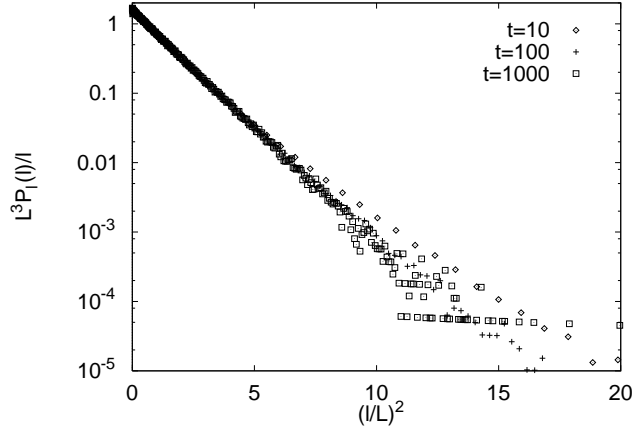}
\caption{Fit of the data in Fig.\ \protect{\ref{fig3}} to the form 
(\protect{\ref{scalepl}}).
\label{fig4}}
\end{figure}

Figure \ref{fig3} is a scaling plot of the domain size distribution 
$P_l(l)$ (defined as the number of domains  of size $l$ per lattice site), 
for a simulation with $\mu=0.5$.  The average domain size $L(t)$ is defined 
by $L^{-1}=\sum_l P_l(l)$.
The data show good collapse to scaling, even for short times.  Figure 
\ref{fig4} shows the same data plotted in the form $\ln[L^3 P_l(l)/l]$ 
versus $(l/L)^2$.  The linear behaviour evident in the plot suggests 
scaling of the form 
\begin{equation}
P_l\propto{l\over L^3(t)} \exp\left\{-\lambda\left[
l/ L(t)\right]^2\right\}\label{scalepl}
\end{equation}
For the case $\mu\ne 0.5$, the average sizes of `up' and `down' domains 
differs by a factor $\mu/(1-\mu)$, but nevertheless the up- and down-domain
size distributions $P_+$ and $P_-$
are both found to satisfy independently scaling of the form 
(\ref{scalepl}).

\begin{figure}
\epsfxsize=\hsize
\epsfbox{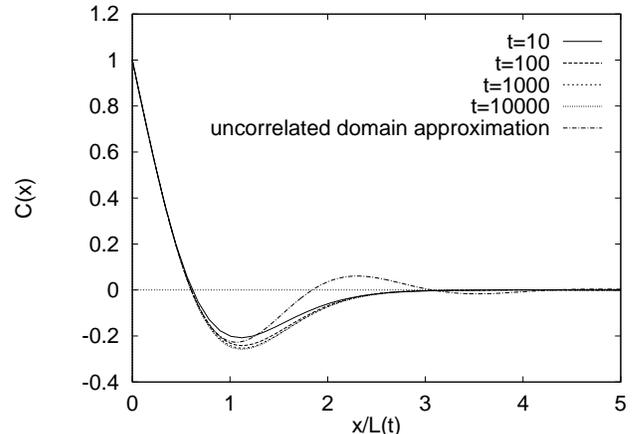}
\caption{Scaling plot of the equal-time spin-spin correlation function for 
$\mu=0.5$.
The prediction assuming uncorrelated domains is shown for comparison.
\label{fig5}}
\end{figure}

Figure \ref{fig5} 
shows the scaling of the equal-time two-point correlation function
$C(x, t)=\langle S_i(t) S_{i+x}(t)\rangle$.  Although the data appear to 
collapse to a scaling form, the approach appears slower than was the case 
for $P_l$, suggesting that $P_l(l)$ is a more natural quantity to 
describe the system.  The apparent simple form for $P_l$ suggests that the 
scaling state might be very simple, for instance there might be no correlations
between domains in the scaling limit.  The structure factor $S(q)$ 
[the Fourier transform of $C(x)$] for a system consisting of uncorrelated 
domains may be shown to be \cite{Kawasaki}
\begin{equation}
S(q)={4\over Lq^2}{1-\left|{\tilde P}_l(q)\right|^2\over
\left|1+{\tilde P}_l(q)\right|^2},
\label{sq}
\end{equation}
where ${\tilde P}_l(q)$ is the Fourier transform of $P_l$.  The inverse Fourier
transform of (\ref{sq}), where ${\tilde P}_l$ was calculated by assuming
the simple form (\ref{scalepl}), is plotted in Fig.\ \ref{fig5} 
for comparison.  The discrepancy with the
simulation data shows that strong correlations need to be taken in to account 
in this system, even though simulations measured only $\sim 3$--$5\%$ 
correlations between the sizes of neighbouring domains.

\subsection{Solution for domain density assuming scaling}

The numerical data suggest that the scaling function for the domain size 
distribution is the same for minority and majority domains, and is independent
 of $\mu$.  Using this as an assumption, we can calculate the average 
domain size.  The number of domain walls changes by a domain of size one 
shrinking to nothing, and so we expect that $P_\pm(l_\pm)\sim l_\pm$ for
$l_\pm\to 0$.  We therefore assume a scaling form 
\begin{equation}
P_\pm(l_\pm, t)=\alpha_\pm {l_\pm \over L_\pm^3(t)} F\left( {l_\pm\over
L_\pm(t)}\right), 
\end{equation}
where $\alpha_\pm$ is chosen so that $F(0)=1$, and $F'(0)=0$.  The rate at 
which `up' domains vanish is $P_+(1)$, which approaches $\partial P_+/
\partial l_+$ in the continuum limit.
Since each domain vanishing event removes 2 domain walls, the rate of 
decay of the number $N(t)$ of domain walls per site is
\begin{equation} 
-{dN\over dt} = 2\left(\left.{\partial P_+\over\partial l_+}\right|_{l_+=0}
+ \left.{\partial P_-\over\partial l_-}\right|_{l_-=0}\right) = 
2\left({\alpha_+\over L_+^3}+{\alpha_-\over L_-^3} \right).\label{eomN}
\end{equation}
The domain size distribution is normalized to the density of domains, and 
there are the same number of `up' as `down' domains, so
\begin{eqnarray}
N/2 &=& \int_0^\infty P_+(l_+)\, dl_+ = {\alpha_+ f_1\over L_+}\label{Nplus}\\
N/2 &=& \int_0^\infty P_+(l_-)\, dl_- = {\alpha_- f_1\over L_-}\label{Nminus}
\end{eqnarray}
where $f_1 \equiv \int_0^\infty xF(x)\, dx$. 
The densities of up and down spins are 
\begin{eqnarray}
\mu &=& \int_0^\infty l_+ P_+(l_+)\, dl_+ = \alpha_+ f_2, \label{muplus}\\
1-\mu &=& \int_0^\infty l_- P_-(l_-)\, dl_- = \alpha_- f_2\label{muminus},
\end{eqnarray}
where $f_2\equiv\int_0^\infty x^2 F(x)\, dx$.  Substituting 
for $L_\pm$ and $\alpha_\pm$ from (\ref{Nplus}--\ref{muminus})
into (\ref{eomN}), and integrating, we find the following asymptotic 
result as $t\to\infty$:
\begin{equation}
N(t)=\left\{ {2 f_1^3\over f_2^2 t\left[\mu^{-2}+(1-\mu)^{-2}\right]}
\right\}^{1/2}
\end{equation}

For the particular case $F(x)=\exp(-\lambda x^2)$, suggested by the data, 
we have $f_1=1/(2\lambda)$, $f_2=\pi^{1/2}\lambda^{-3/2}/4$, and
\begin{equation}
N(t)={2\over \left\{\pi t\left[\mu^{-2}+(1-\mu)^{-2}\right]\right\}^{1/2}}.
\label{resultN}
\end{equation}
The straight lines in Fig.~\ref{fig1} are, in fact, plots of equation 
(\ref{resultN}) for appropriate values of $\mu$.
The excellent agreement of the data with the prediction confirms both the 
predicted $\mu$-dependence and also the simple form for $F(x)$.

\subsection{Solution for $P_\pm$ in the limit $\mu\to 0$.}
The simulations are in excellent agreement with the scaling hypothesis, and 
with the simple form for the scaling function (\ref{scalepl}).  We would like 
to have an {\it ab initio\/} explanation for these results.  Unfortunately, we
were only able to solve the dynamics in the `off-critical' limit $\mu\to 0$ 
[or, equivalently, $\mu\to 1$].

In the limit $\mu\to 0$, it is necessary only to consider the motion of the 
minority spins, i.e.\ the motion of the majority domains. This is because the 
size of each domain performs a random walk until it either dies (shrinks to 
zero size) or coalesces with the nearest domain of the same type.  To 
coalesce with another domain, the intervening domain of the opposite type 
has to shrink to zero size.  The time scale  for the vanishing of
majority domains
will therefore be much longer (by a factor $\sim\mu^{-2}$) than for  minority 
domains.
The dynamics will therefore progress primarily by minority domains shrinking 
to zero size, and never coalescing, whereas the majority domains' sizes only 
change appreciably due to coalescence.

Consider a particular minority domain containing $n$ `up' spins at time $t$.
In the limit where coalescence is forbidden, the domain changes size by:
(a) an `up' spin arriving from the next domain to the left; or (b) an `up'
spin splitting off the domain, and moving to the next domain on the right.
It is also possible for a `down' spin to move from the right to the left
of the domain, but this does not change the value of $n$.   Notice that the 
dynamics is independent of where the neighbouring domains are, whether they 
are vanishing or coalescing. Once the size of the domain reduces to zero, 
the domain ceases to exist. The Master equation for the probability 
$P(n,t)$ of the size being $n$ at time $t$ is
\begin{equation}
{dP(n, t)\over dt} = P(n+1, t)+P(n-1, t)-2P(n, t)
\label{mepn}
\end{equation}
for $n\ge 1$, with $P(0, t)=0$. In the continuum limit this Master 
equation approaches the diffusion equation , whose solution with 
$P(n,t=0)=\delta(n-n_0)$ is
\begin{eqnarray}
P(n,t)&=&\left(4\pi t\right)^{-1/2}\left\{
\exp\left[-{\left(n-n_0\right)^2\over 4t}\right]\right. \nonumber\\
&\phantom{=}& \left.\phantom{
\left(4\pi t\right)^{-1/2}\left\{\right.
}
-
\exp\left[-{\left(n+n_0\right)^2\over 4t}\right]\right\} \\
&=&\left(\pi t\right)^{-1/2}\sinh \left({nn_0\over 2t}\right)\,\exp\left(
-{n^2\over 4t}-{n_0^2\over 4t}\right).
\end{eqnarray}
In the scaling limit $t\to\infty$, $n\to\infty$, with $n/\sqrt{t}$ fixed,
this reduces to
\begin{equation}
P(n,t)\to {nn_0\over 2\sqrt{\pi t^3}}\exp\left(-{n^2\over 4t}\right).
\end{equation}
In a random initial state, the number of domains of size $n_0$ per lattice site
is $(1-\mu)^2\mu^{n_0}$.  The `up' domain size distribution at a time $t$
in the scaling limit is therefore
\begin{eqnarray}
P_+(l_+)&=&\sum_{n_0}(1-\mu)^2\mu^{n_0}P(l_+,t)\\
&\to&{\mu l_+\over 2\sqrt{\pi t^3}}\exp\left(-{l_+^2\over 4t}\right),
\end{eqnarray}
which is of the form (\ref{scalepl}), with
$L=2t^{1/2}$. The total domain density (twice the density of `up' domains) is
\begin{equation}
N(t)=\int_0^\infty P_+(l_+)\,dl_+ = {2\mu\over\sqrt{\pi t}},
\end{equation}
which approaches (\ref{resultN}) in the limit $\mu\to 0$.

We may calculate the size distribution of the majority domains from the 
probability that a given region contain no domain walls.  Consider a 
region of $M$ lattice sites containing $n$  minority spins; these 
spins need not necessarily all belong to the same domain.  Then the number of 
minority spins changes by spins entering the region at the left and 
coalescing with the leftmost domain wall in the region, and spins splitting 
off the rightmost domain wall.  Once $n$ reduces to zero, however, any spins 
entering the region from the left will simply pass through the region, and 
it will remain empty.  We have again assumed the limit $\mu\to 0$, by not 
allowing for any domains to move out of the region or to coalesce with 
domains outside the region. 

Under these conditions, the probability $P_M(n, t)$ obeys a Master 
equation of the form (\ref{mepn}), whose solution with $n=\mu M$ at $t=0$ is
\begin{eqnarray}
P_M(n,t)&=&\left(4\pi t\right)^{-1/2}\left\{
\exp\left[-{\left(n-\mu M\right)^2\over 4t}\right]\right. \nonumber\\
&\phantom{=}& \left.\phantom{
\left(4\pi t\right)^{-1/2}\left\{\right.
}
-
\exp\left[-{\left(n+\mu M\right)^2\over 4t}\right]\right\}
\end{eqnarray}
The probability $P_E$ of the region $M$ being empty is then 
\begin{eqnarray}
P_E(M,t)&=&1-\int_0^\infty P_M(n, t)\, dn\\
& = &1-(4\pi t)^{-1/2}
\int_{-\mu M}^{\mu M}\exp\left(-{u^2\over 4 t}\right)\, du
\end{eqnarray}
Consider now a region of $m$ sites, sitting inside a domain of size 
$l$ sites.  The number of positions that it can occupy within the region 
is $(l-m)\Theta(l-m)$, where $\Theta$ is the Heavyside function.  The 
probability that a randomly-chosen interval $m$ lies within a domain of 
size in the range $l$ to $l+dl$ is therefore $(l-m)\Theta(l-m)P_l(l)\, dl$, 
so the probability that a region of size $m$ contains no domain walls is
\begin{equation}
P_E(m)=\int_m^\infty (l-m)P_l(l)\, dl.
\end{equation}
Differentiating twice, we have $P_E''(m)=P_l(m)$. 

The result for the distribution of majority domains is therefore
\begin{eqnarray}
P_-(l_-)&=&\left.{\partial^2 P_E(M)\over \partial M^2}\right|_{M=l_-}\\
&=&{\mu^3 l_-\over 2 \sqrt{\pi t^3}}\exp\left(-{\mu^2l_-^2\over 4t}\right),
\end{eqnarray} 
which is of the form (\ref{scalepl}) with $L_-=L_+/\mu$.  This shows that the
calculation is only valid to lowest order in $\mu$, since the conservation of 
magnetisation implies that  $L_-/L_+=(1-\mu)/\mu$.

\section{Persistence exponent}
It was recently found that, in a 1D Ising model evolving at zero temperature 
from a disordered state under Glauber dynamics, the probability that a given 
spin has never flipped up to time $t$ decays as $t^{-\theta}$ 
\cite{persistence}. The value of the `persistence exponent' $\theta$ depends 
upon the magnetisation, and whether the spin is in the majority or minority 
phase. The dynamics for a spin in a phase with volume fraction $\mu$ is the 
same as for a $q$-state Potts model with symmetric initial condition, with 
$q=1/\mu$ \cite{Boston}, leading to the result that $\theta$ takes values 
in the range 0--1 as $\mu$ decreases from 1 to 0, with $\theta=3/8$ for 
$\mu=1/2$ \cite{DHP}.

For a 1D driven diffusive system, there are two kinds of persistence that 
may be considered.  The first concerns the probability that a spin in the 
microscopic Ising representation never having changed its value.  Whenever 
a domain wall emits a spin, that spin moves rapidly through a domain, 
causing each of the spins in that domain to have flipped twice. Since this 
spin emission is a Markov process, the probability that a given spin has 
never flipped decays exponentially.

A more interesting kind of persistence to investigate is the probability that 
a given spin has never belonged to another {\sl stable} domain.  That is, we 
discount the rapid flipping due to spin motion through a domain, and consider
only the case where an entire domain has migrated towards the site in question.
This kind of coarse-graining in time would also be necessary when studying 
the Glauber-Ising model at low but non-zero temperatures, where short-lived 
thermally-activated flipping of spins within the interiors of domains occurs, 
in order to recover the true zero-temperature persistence behaviour.

\subsection{Analytical results for $\mu\to 0$}
Let us consider a test site initially within a minority domain, a distance 
$n_1$ sites from the left domain wall and $n_2$ sites from the right wall.  
We define $n_1$ and $n_2$ such that a spin in a domain of size unity has 
$(n_1,n_2)=(1,1)$.  The dynamics causes the domain walls to wander
stochastically, and eventually one of the walls will cross the test site, 
i.e.\ the test site will have flipped.

We shall assume that this domain does not coalesce with another domain before 
one of the domain walls reaches the test site. This assumption will certainly 
be valid in the limit $\mu\to 0$. Then the three processes that cause the 
domain walls to move are: (i) an `up' spin joins onto the left-hand edge, 
$(n_1, n_2)\to (n_1+1, n_2)$; 
(ii) an `up' spin splits off the right-hand edge, 
$(n_1, n_2)\to (n_1, n_2-1)$;
(iii) a `down' spin splits off the right-hand edge and moves through the 
domain to the left-hand edge, $(n_1, n_2)\to (n_1-1, n_2+1)$.  The Master 
equation for the joint probability $P(n_1, n_2, t)$ is then
\begin{eqnarray}
{dP(n_1, n_2, t)\over dt} = 
P(n_1-1, n_2)+P(n_1+1, n_2-1)&&\\
 + P(n_1, n_2+1) - 3P(n_1, n_2)&&
\end{eqnarray}
which becomes, in the continuum limit $n_1\to x_1$, $n_2\to x_2$,
\begin{equation}
{\partial P(x_1, x_2, t)\over\partial t}
=\left\{ {\partial^2\over\partial x_1^2}+{\partial^2\over\partial x_2^2}
-{\partial^2\over\partial x_1\partial x_2}\right\}P.\label{eomnn}
\end{equation}
Making the change of variable
\begin{eqnarray}
x&=&x_1+x_2\\
y&=&{1\over\sqrt{3}}\left(x_1-x_2\right),
\end{eqnarray}
equation (\ref{eomnn}) becomes
\begin{equation}
{\partial P(x, y, t)\over\partial t}
=\left\{ {\partial^2\over\partial x^2}+{\partial^2\over\partial y^2}
\right\}P.\label{eomxy}
\end{equation}
When either of the domain walls reaches the test site, the test site flips.
Therefore, if we want $P$ to represent the conditional probability that the
site has not flipped, we must solve equation (\ref{eomnn}) with boundary
condition $P=0$ along $x_1=0$ and $x_2=0$.  This corresponds to solving the 
diffusion equation (\ref{eomxy}) with boundary condition $P=0$ along
the lines $y=\pm x/\sqrt{3}$, for the region $x\ge y\sqrt{3}$, $x\ge 
-y\sqrt{3}$---that is, in a wedge of angle $\pi/3$ with absorbing boundaries.

The diffusion equation is readily solved in a wedge of angle $\psi$
with absorbing boundaries, and it is known that the survival probability
for a random walker under such conditions decays as $t^{-\pi/2\psi}$ 
\cite{CJF}. After transforming equation (\ref{eomxy}) into polar 
coordinates $r$ and $\theta$, where
\begin{eqnarray}
r&=&\left(x^2+y^2\right)^{1/2}\\
\theta&=&\phi+\pi/6\\
\tan\phi&=&y/x,
\end{eqnarray}
we need to solve the diffusion equation for the region $0\le\theta\le\pi/3$,
with $P=0$ on $\theta=0$ and $\theta=\pi/3$.  The appropriate solution, 
starting from $r=r_0$, $\theta=\theta_0$ is
\begin{eqnarray}
P(r, \theta, t)&=&{3\over\pi}\sum_m\int_0^\infty{d\lambda\, e^{-\lambda t}}
\sin (3m\theta)\sin(3m\theta_0)\times \nonumber\\
&&\phantom{{3\over\pi}
\sum_m\int_0^\infty
}
\times J_{3m}(r_0\lambda^{1/2})J_{3m}(r\lambda^{1/2})\\
&=&{3\over\pi t}\sum_m\sin(3m\theta)\sin(3m\theta_0)\times\nonumber\\
&&\phantom{
{3\over\pi t}\sum_m
}\times
I_{3m}\left({rr_0\over2t}\right) \exp\left(-{r_0^2+r^2\over 4t}\right)
\end{eqnarray}
where we have used eqn.~6.615 of \cite{Gradshteyn}.  To find the persistence 
probability, we need to evaluate $P_P(t)=
\int_0^\infty r\,dr\int_0^{\pi/3}d\theta\,P$.  
Performing the integrals \cite{Gradshteyn}, and taking the limit 
$t\to\infty$ (where the dominant contribution comes from the term
$m=1$ in the sum)
we find
\begin{equation}
P_P\to{1\over 2\sqrt{\pi}}
\sin (3\theta_0)\left({r_0^2\over 4t}\right)^{3/2}.\label{mess}
\end{equation}
On substituting for the initial conditions in terms of $x=n_1^0+n_2^0$, 
$y=(n_1^0-n_2^0)/\sqrt{3}$,
equation (\ref{mess}) reduces to
$P_P(t)={1\over 4\sqrt{\pi}}n_1^0n_2^0(n_1^0+n_2^0)t^{-3/2}$.

In order to calculate the persistence probability, we need to average over 
the possible values of $n_1^0$ and $n_2^0$.  The probability in the initial 
state of a given `up' spin having precisely $(n_1^0-1)$ consecutive `up' 
neighbours to its left is $(1-\mu)\mu^{n_1^0-1}$, so the probability of an 
initial configuration $(n_1^0, n_2^0)$ is $(1-\mu)^{2} \mu^{n_1^0+n_2^0-2}$. 
Summing over $n_1^0\ge 1$ and $n_2^0\ge 1$ gives the final result,  
\begin{equation}
P_P(t)={1\over 2\sqrt{\pi t^3}}{ 1+\mu\over (1-\mu)^3}.\label{pp}
\end{equation}

One might attempt to perform a similar calculation for a site within 
a majority domain.  Here, however, it is important to take account not only
of the fact that the walls of the domain can move, but also that the 
neighbouring domains can vanish before the test spin has flipped.
Contributions from domains that have coalesced will therefore be important, 
and the necessity to include information about domain-domain 
correlations makes the calculation extremely complicated, if 
not intractable.

\subsection{Simulation results}
The persistence probability for both majority and minority spins was 
measured for a range of values of $\mu$.  While the measured values of the 
exponent $\alpha$ for the minority domains was found to be close to $1.5$ 
for $\mu$ small, for $\mu=0.5$ there appeared to be some evidence that the 
asymptotic behaviour was governed by a different exponent.  However, the 
slow onset of the asymptotic regime for intermediate values of $\mu$, 
together with the large value of the exponent ($> 1$, compared with $\le 1$ 
for the Glauber case \cite{DHP}), led to unavoidably poor statistics in the 
asymptotic regime, and hence it was not possible to establish reliable values 
for the exponent.

\begin{figure}
\epsfxsize=\hsize
\epsfbox{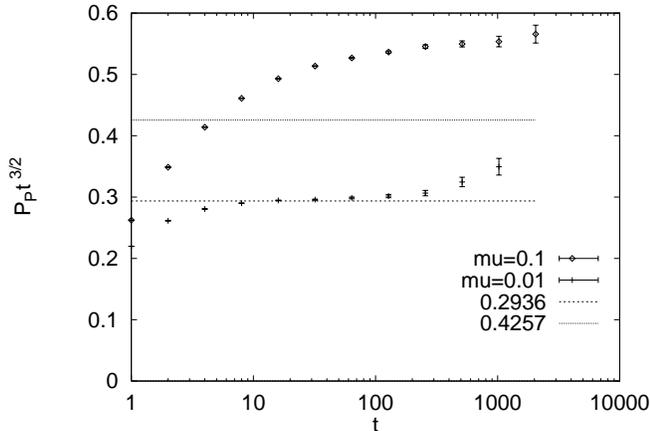}
\caption{Bias plot for the persistence probability of minority spins. 
\label{fig6}}
\end{figure}

Figure \ref{fig6} 
shows a plot of $P_P(t)t^{3/2}$ against time for minority spins, 
for $\mu=0.01$.  For comparison, the constant ($\approx 0.2936$)
predicted by equation (\ref{pp}) is also shown.  There is reasonable 
agreement of the simulations with the independent-domain prediction, 
suggesting that it is a good approximation for small $\mu$, for the time 
regime measured.  We interpret the deviation from the constant at longer 
times as more likely to be due to statistics than a true systematic effect.
Similar data for $\mu=0.1$ is shown, and deviations from the constant 
predicted by (\ref{pp}) ($\approx 0.4257$) are already quite marked.

\begin{figure}
\epsfxsize=\hsize
\epsfbox{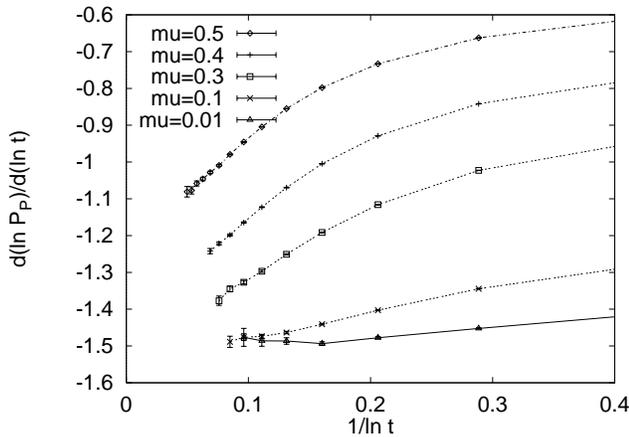}
\caption{Effective exponent for the persistence probability of minority spins.
\label{fig7}}
\end{figure}

The effective persistence exponent $d(\ln P_P)/d(\ln t)$ for
minority spins is plotted as a function of $1/\ln t$ in Figure \ref{fig7}, 
for several values of $\mu$.  The purpose of this plot is to investigate 
whether there are any underlying trends in the exponent.  When a function 
of the form $f(t)\propto t^a(\ln t)^b$ is plotted on a graph of this kind, 
the curve approaches the $f$-axis linearly with gradient $b$ and intercept 
$a$.  The behaviour for data that converge asymptotically to a power-law  
is for the slope of such a plot to level off to zero in the asymptotic regime.
For small values of $\mu$, the exponent seems to have settled down to a value 
close to $-1.5$.  However, for $\mu$ close to $0.5$ there still appears to be 
some trend in the effective exponent, suggesting that the asymptotic regime 
has not been reached.  Only for $\mu=0.5$ is there strong evidence that the 
data converge to an exponent different from $-1.5$. Improved statistics and 
longer times would be needed to give an unequivocal conclusion, but we would 
not be able to do this using our current techniques since each set of data 
required approximately 1 week of CPU time.

The results for the persistence of the majority species are even more 
problematical, because for $\mu\ne 0.5$ too few spins were found to have 
to flipped for the asymptotic regime to have been reached.

The deviations from the free-domain picture must be due to the fact that some 
of the minority domains coalesce.  It is possible to estimate the number of 
minority domains that coalesce from the rate at which majority domains vanish.
From equation(\ref{eomN}), using
(\ref{Nplus}--\ref{muminus}), we find
\begin{equation}
-{dN_+\over dt}={(1-\mu)^2+\mu^2\over \mu^2}
\left.{\partial P_-\over\partial l_-}\right|_{l_-=0},
\end{equation}
where $N_+=N/2$ is the number of minority (`up') domains per site.  
Integrating, we find
\begin{equation}
 \int_t^\infty dt\left.{\partial P_-\over\partial l_-}\right|_{l_-=0} 
={\mu^2\over (1-\mu)^2 + \mu^2} N_+(t).
\end{equation}
The term on the left-hand side is the total number of majority-domains
that vanish between time $t$ and infinity, which is equal to the number of 
minority-domain coalescences.  Some domains may coalesce more than once,
so this is an upper bound on the number of minority domains that
coalesce at least once after time $t$.  The fraction of minority domains 
that coalesce after time $t$ is therefore less than $\mu^2/\left[\mu^2
+(1-\mu)^2\right]$, which vanishes like $\mu^2$ for small $\mu$.

Measurements from the numerical simulations confirmed the picture that 
the fraction of domains that coalesce is small for $\mu$ small.  
Nevertheless, it is found that the dominant contribution to persistence 
in the long-time limit comes from spins in domains that have coalesced, 
thus explaining the deviations from the free-domain approximation.

\section{Summary}
The low-temperature coarsening dynamics of a driven diffusive system -- the 
1-D Ising model with a driving force $E$ satisfying $T \ll E \ll J$ -- has 
been studied by a combination of analytical and numerical techniques. 
Compelling evidence for a mean domain size growing as $t^{1/2}$, and a 
domain-size distribution of the form (\ref{scalepl}), has been presented. 
These results are exact in the limit where one phase occupies a vanishingly 
small volume fraction $\mu$. In the same limit, the persistence exponent 
for the minority phase is $\theta =3/2$.  The limit of
zero volume fraction was studied  by Lifshitz and Slyozov \cite{ls}  
to predict the 
growth exponent ($=1/3$) for the case without driving force in 
general dimension, and it is hoped that 
the approach of the present paper might also be usefully extended
to higher dimensions.

The random-walk character of the domain dynamics suggests $t^{1/2}$ growth 
generally, and this is borne out by the simulation results (Figures 1 and 2). 
The simulations also lend strong support (Figures 3 and 4) to the scaling 
distribution (\ref{scalepl}) for general $\mu$. By contrast, the value of 
$\theta$ away from the small-$\mu$ limit is difficult to determine 
numerically, due to slow convergence to the asymptotic regime (Figure 7). 
For $\mu=0.5$, however, the results seem to be inconsistent with 
$\theta=3/2$, suggesting that the exponent $\theta$ may depend continuously 
on $\mu$, as is the case for the 1-D Ising model with Glauber Dynamics 
\cite{Boston}. 

It is tempting to propose an experimental setup that might be described by
the present one-dimensional model.  If two immiscible
fluids of differing density
are stirred and placed in a vertical tube, then they will typically be able to 
slide past each other and separate hydrodynamically.  However, if the 
tube is sufficiently narrow then 
it is possible for a state where the denser fluid is above the lighter fluid
to be metastable;  the loss in gravitational 
potential energy if the  interface 
is tilted slightly can be stabilised by the increase in interfacial energy
as the area of the interface increases.  A state consisting of 
alternating quasi-one-dimensional `domains' can therefore be metastable.
There now arises the question of adding a noise source to create droplets
at the surface.   Simply shaking the tube will tend to induce 
tilting of the interface, which will lead to hydrodynamic instability.
The driving vibrations therefore have to be of a wavelength much
smaller than the width of the tube in order to create droplets without
exciting the `sloshing' mode.  We leave the problems of finding an 
ultrasonic source of high enough intensity, and of preparing the intial 
condition, as challenges for the keen experimenter.

In a separate paper \cite{Caroline} we will present results for the $T=0$ 
dynamics of a deterministic 1-D scalar field model, defined by the modified 
Cahn-Hilliard equation 
$\partial_t\phi = -\partial_x^2(\partial_x^2\phi +\phi-\phi^3) + 
E\phi\partial_x\phi$. In the small-$E$ limit, the coarsening dynamics of 
this driven diffusive model also exhibit a scaling distribution for domain 
sizes, and a $t^{1/2}$ growth of the mean domain size. Both models, the 
stochastic Ising model considered here, and the deterministic model are of 
great interest in higher dimensions. The approach of looking at the 
limit of small volume fraction $\mu$, which proved so successful here, 
may well be fruitful in elucidating the behaviour of these models in 
general dimension $D$.

\end{multicols}

\end{document}